# Optical loss by surface transfer doping in silicon waveguides

L. Alloatti[1,2,a)], C. Koos[2,3], J. Leuthold[2,3,4]

[1] Massachusetts Institute of Technology, 77 Massachusetts Ave., Cambridge, MA 02139, USA
[2] Institute of Photonics and Quantum Electronics (IPQ) and Institute of Microstructure Technology (IMT), Karlsruhe Institute of Technology (KIT), 76131 Karlsruhe, Germany
[3] Physikalisches Institut and DFG-Center for Functional Nanostructures, Karlsruhe Institute of Technology, P.O. Box 6980, 76049 Karlsruhe, Germany
[4] Institute of Electromagnetic Fields (IEF), ETH Zurich, Zurich, Switzerland

We show that undoped silicon waveguides may suffer of up to 1.8 dB/cm free-carrier absorption caused by improper surface passivation. To verify the effects of free-carriers we apply a gate field to the waveguides. Smallest losses correspond to higher electrical sheet resistances and are generally obtained with non-zero gate fields. The presence of free carriers for zero gate field is attributed to surface transfer doping. These results open new perspectives for minimizing propagation losses in silicon waveguides and for obtaining low-loss and highly conductive silicon films without applying a gate voltage.

©2015 American Institute of Physics

Low optical propagation losses in silicon waveguides are key for minimizing the total power budget of optical transceivers,[1] guaranteeing high nonlinear conversion efficiencies[2] or minimizing parasitic perturbations of optical quantum states.[3] While the smallest values reported so far are 0.27 dB/cm in amorphous-silicon rib waveguides,[4] and 0.45 dB/cm for fully-etched waveguides in crystalline silicon,[5] the ultimate propagation loss limits of bulk crystalline silicon remain elusive.[6]

Several causes of optical loss in silicon waveguides have already been identified to date: surface roughness,[7,8,9,10] substrate leakage,[7] absorption from surface-states,[11] absorption from vibrational states[12] and free-carrier-absorption.[13,14,15] More specifically, by performing iteratively chemical cleaning procedures, the improved propagation loss in narrow waveguides has been attributed to a decreased side-wall roughness,[9] however, experiments on high-Q resonators have shown that surface passivation may play a dominant role when compared to the latter.[11] Among all loss mechanisms mentioned above, free-carrier absorption plays an important role, especially when it comes to electro-optical devices such as modulators, where the silicon must exhibit at the same time high electrical conductivity and small optical loss. Free-carrier absorption in bulk doped silicon can be described by the equations published by Soref et al..[13,14] However these equations need to be corrected if the carriers originate from an electrostatic gate field rather than from ion implantation.[15]

In this work we identify an additional cause of optical absorption which can lead to propagation losses as high as 1.8 dB/cm in standard silicon waveguides. The loss is attributed to the presence of free carriers caused by the interaction of the silicon waveguide with the surface. This effect has recently been reported to modify the sheet resistance of thin silicon membranes and was named surface transfer doping.[16,17] While we confirm that chemical cleaning has beneficial effects on the propagation losses as shown by Sparacin et al.,[9] we attribute this effect to reduced surface transfer doping and hence reduced free-carrier absorption. Our findings are consistent also with the experimental results on surface passivation of Borselli et al.[11] and may provide an alternative interpretation for the loss reduction observed in waveguides coated by atomic layer deposition (ALD).[18]

To identify the effects of different passivation processes on the optical loss, we use rib waveguides defined in the device layer of silicon-on-insulator (SOI) wafers from Soitec, Fig. 1. The device layer is 220 nm thick, the buried-oxide (BOX) is 2 $\mu$m high, the rib is 700 nm wide, and the partial etch is of 70 nm. The residual p-type impurity concentration of $N_A \approx 10^{-15}\,\mathrm{cm}^{-3}$ deduced from the Soitec specifications would account for a bulk free-carrier propagation loss of less than 0.03 dB/cm.[13] The fundamental quasi-TE$_{00}$ mode is concentrated in the waveguide core and the optical transmission is recorded as a function of a gate voltage $V_\mathrm{gate}$ applied between the silicon substrate and the grounded silicon waveguide, Fig. 1. By recording the optical transmission for different waveguide lengths,[15] we could extract the propagation loss, Fig. 2.

The fabricated chips were treated with different cleaning procedures, and coated with different materials, as summarized in Table 1. For sample 5, ultra-clean chemicals were used and a full RCA cleaning was performed.[19] In the other cases, no particular attention to contaminations was paid, and an organic cleaning sequence with acetone, isopropanol and water was performed in cleanroom conditions, Table 1.

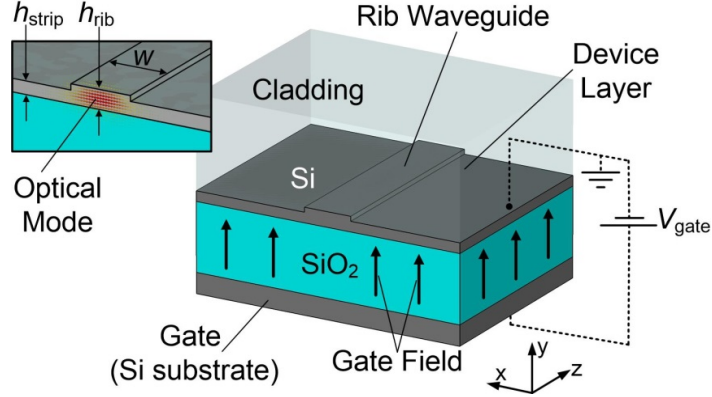

Fig. 1. Cross-section of the silicon waveguide used for optical loss vs. gate field measurements. The waveguide is located on top of a low-refractive index buried oxide layer ($SiO_2$). The optical quasi-$TE_{00}$ mode is confined in the upper undoped silicon layer (device layer). A gate voltage ($V_{gate}$) is applied between the silicon substrate and the grounded device layer therefore changing the carrier density in the waveguide and the optical loss. The rib height is $h_{rib} = 220$ nm, the strip height is $h_{strip} = 150$ nm, the rib width is $w = 700$ nm, and the silicon-oxide thickness is $d_{SiO_2} = 2\,\mu m$.

The optical loss vs. gate field is shown for three representative samples in Fig. 2. For all samples, the optical loss was reduced by applying a negative gate field. For the PMMA-coated waveguide in Fig. 2 the zero-field excess loss is of 1.8 dB/cm. The zero-field excess loss for the other waveguides is summarized in Table 1. We further investigated the resistance of the samples as a function of the gate field and observed that when the optical loss is minimum the sheet resistance is maximum, similarly to what we observed previously.[15] In particular, for sample 1 we achieved a sheet resistance of $\approx 10\,k\Omega$/sq without applying a gate field. The excess optical loss observed when no field is applied is therefore attributed to the presence of charges and related free-carrier absorption. Since the silicon did not undergo any high-temperature step which may cause diffusion of impurities, and since the loss is found to be a sensitive function of the passivation process, the loss is attributed to surface transfer doping.[16,17]

| Sample ID | Fabrication | Cleaning | Cladding | Zero-field excess loss [dB/cm] |
|---|---|---|---|---|
| 1 | IMEC | ACE, ISO, DI | PMMA | 1.8 |
| 2 | IMEC | ACE, ISO, DI | M3 | 0.7 |
| 3 | IMEC | As shipped | As shipped | 1.6 |
| 4 | In-house | ACE, ISO, DI | PMMA | 0.5 |
| 5 | In-house | Ultra-clean RCA | PMMA | 0.28 |

Table 1. Tested samples. The sample identity (ID) is defined for later reference. The fabrication occurred either at IMEC or in-house.[15] Two cleaning procedures were used: standard organic cleaning comprising acetone (ACE), isopropanol (ISO) and deionized water (DI) rinse or standard RCA-cleaning with ultrapure chemicals as described previously.[15] Different claddings were used: PMMA 950k,[15] the organic nonlinear cladding M3,[20] and a standard photoresist deposited by IMEC for protecting the waveguides during shipping. The excess loss refers to the extra loss at zero gate voltage compared to the minimum attainable. For all samples the minimum loss is observed for negative gate fields (negative voltage on the silicon substrate).

Two independent sets of samples were investigated. One set was produced by IMEC, and the other was fabricated in-house by e-beam lithography, Table 1. Samples 4 and 5 correspond to the same chip measured before and after the RCA cleaning step with ultra-clean chemicals.

We have further investigated this effect in crystalline strip waveguides obtained in the IBM 45 nm 12SOI process.[21] To this end, we have fabricated two optical waveguides with identical geometry (same grating couplers, same number and type of bends, strip width of 470 nm), with 0.1 cm and 1.1 cm of length. Each waveguide was contacted electrically with a 100 nm thick silicon strip touching one of the grating couplers, and a wire located 2.2 $\mu$m above the silicon surface was acting like a gate. The optical loss was tested at 1200 nm as a function of the gate voltage. The optical loss could not be diminished by applying a gate voltage, therefore indicated that surface charge transfer does not occur in this process.

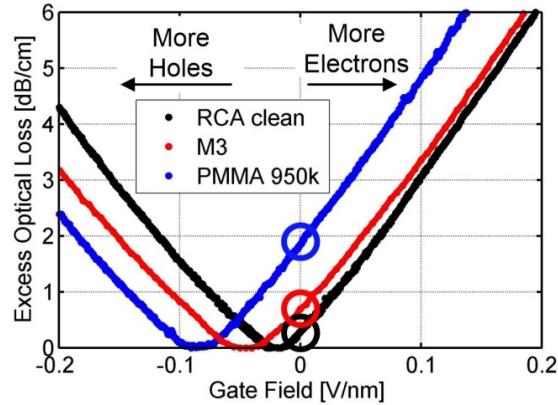

Fig. 2. Measured excess optical loss for samples with different surface treatments (sample 1, 2 and 5) as a function of the applied gate field. For large positive (negative) gate fields, an electron (hole) layer is formed, and the optical loss increases. The circles indicate the extra-loss present in normal conditions (when no gate field is applied).

In our case we do not know which constituent or mechanism is acting as a surface transfer dopant. In fact, contamination may originate from the reactive-ion etching (RIE) process or may be present in the chemicals utilized for cleaning or coating the waveguides. Irrespective of the exact source of the contamination, this work demonstrates that carriers originating from surface transfer can cause significant optical absorption in silicon waveguides. Secondly, it suggests that surface transfer doping may cause undesired carrier depletion of thin doped silicon slabs – an effect which may explain the high resistances obtained in different doped devices.[20,22] Finally, this work opens novel perspectives, such as the possibility of achieving small sheet resistances and small optical losses[15] by injecting charges into the silicon by utilizing a suited "surface dopant" instead than by applying a gate voltage.[23] This may enable gateless modulators with bandwidths of 100 GHz and beyond.[20]

In conclusion we have shown that the optical loss of standard undoped silicon waveguides may be reduced by 1.8 dB/cm by applying a gate field. Lower losses correspond to higher sheet resistances and therefore to lower free-carrier densities inside the waveguide. The presence of free-carriers is attributed to surface transfer doping and the excess optical loss to free-carrier absorption. Surface cleaning and passivation based on the RCA process using ultra-clean chemicals showed the smallest zero-field excess optical loss. Surface transfer doping opens new possibilities for obtaining low-loss and highly conductive silicon films without applying an external gate voltage.[20]

We acknowledge Rajeev Ram (MIT) for access to 45nm 12SOI waveguides, and we acknowledge support by the DFG Center for Functional Nanostructures (CFN), the Helmholtz International Research School of Teratronics (HIRST), the Karlsruhe School of Optics & Photonics (KSOP), the EU-FP7 projects SOFI (grant 248609) and PHOXTROT, the BMBF joint project MISTRAL and the European Research Council (ERC Starting Grant 'EnTeraPIC', number 280145). We are grateful for technological support by the Karlsruhe Nano-Micro Facility (KNMF) and the Initiative and Networking Fund of the Helmholtz Association. We further acknowledge the Alfried Krupp von Bohlen und Halbach Foundation and the Deutsche Forschungsgemeinschaft.

a)Electronic mail: alloatti@mit.edu